\documentclass[12pt,preprint]{aastex}

\shorttitle{Squared visibility estimators}
\usepackage{color}
\definecolor{red}{rgb}{1,0,0}
\definecolor{blue}{rgb}{0,0,1}
\newcommand{\shah}{\bigsqcup\!\!\bigsqcup}

\newcommand{\var}{\mathrm{var}}
\newcommand{\covar}{\mathrm{covar}}
\newcommand{\e}{\mathrm{e}}
\begin{document}

\title{Squared visibility estimators \\ Calibrating biases to reach very high dynamic range}

\author{Guy Perrin}
\affil{Observatoire de Paris, LESIA, 92190 Meudon, France}
\email{guy.perrin@obspm.fr}

\author{Stephen T. Ridgway}
\affil{National Optical Astronomy Observatories, Tucson, AZ 85719, USA}
\email{ridgway@noao.edu}

\begin{abstract}

In the near infrared where detectors are limited by read-out noise, most interferometers have been operated in wide band in order to benefit from larger photon rates. We analyze in this paper the biases caused by instrumental and turbulent effects to $V^2$ estimators for both narrow and wide band cases. 
Visibilities are estimated from samples of the interferogram using two different estimators, $V^{2}_1$ which is the classical sum of the squared modulus of Fourier components and a new estimator $V^{2}_2$ for which complex Fourier components are summed prior to taking the square. We present an approach for systematically evaluating the performance and limits of each estimator, and to optimizing observing parameters for each. We include the effects of spectral bandwidth, chromatic dispersion, scan length, and differential piston. We also establish the expression of the Signal-to-Noise Ratio of the two estimators with respect to detector and photon noise. The $V^{2}_1$ estimator is insensitive to dispersion and is always more sensitive than the  $V^{2}_2$  estimator. However, the latter allows to reach better accuracies when detection is differential piston noise limited.  
Biases and noise directly impact the dynamic range of reconstructed images. Very high dynamic ranges are required for direct exoplanet detection by interferometric techniques thus requiring estimators to be bias-free or biases to be accurately calibrated. We discuss which estimator and which conditions are optimum for astronomical applications especially when high accuracy visibilities are required. We show that there is no theoretical limit to measuring visibilities with accuracies as good as $10^{-5}$ which is important in the prospect of detecting faint exoplanets with interferometers. 

\end{abstract}



\section{Introduction}
Optical-infrared interferometers, are providing very high angular resolutions of order 1~mas, a limit to be improved by the next generation instruments by a factor of 10. Several astrophysical topics topics require high dynamic range observations, one of which being the detection of exoplanets around their parent star. The required dynamic range scales with the ratio between the star and the planet, from a few 1000 for hot Jupiters to $10^{10}$ for Earth-like planets in the near-infrared. The dynamic range in an image reconstructed from visibilities and phases can be approximated by the formula of \cite{baldwin2002}:
\begin{equation}
\mathrm{Dynamic}\;\mathrm{range} = \sqrt{\frac{N_{vis}}{(\delta V / V)^2+(\delta \phi)^2}}
\label{eq:dynamique}
\end{equation}
 where $V$ is the visibility modulus and $\delta V$ the associated error, $\delta \phi$ is the phase error of the complex visibility and $N_{vis}$ the number of visibility points. Very large dynamic ranges are achieved with radio interferometers such as the VLA \citep[120 000 for example in ][]{hardcastle2003}. However, infrared interferometers have a very limited number of telescopes with respect to radio and millimetric interferometers thus limiting the number of available visibilities. Eq.~(\ref{eq:dynamique}) shows that large dynamic ranges can only be achieved if visibilities are measured very accurately in this case. Other techniques to detect planets do not necessarily require images. \citet{foresto1997a} have shown that hot Jupiters could be detected by fitting a model to  visibilities with absolute accuracies of $10^{-3}$. Future interferometer facilities will have a sufficient number of telescopes to make better images. But for observations which require the highest dynamic range or for which high photometric precision is mandatory, model fitting will be the most rigourous technique. This process is sensitive to biases. At a more modest level of precision, possible biases must be considered even at the level of $10^{-2}$ when measuring visibilities which change substantially through the bandpass, e.g. spanning visibility nulls. \\

Data reduction procedures must therefore focus on the use of estimators free from the well identified biases such as photon or detector noise. Optical-infrared visibilities are corrupted by biases induced by atmospheric turbulence. Atmospheric turbulence causes coherence losses which are difficult to calibrate because turbulence is not a stationary process and may vary between the source and the calibrator acquisitions. This important issue has been tackled by spatially filtering the beams using single-mode fiber optics, for example. Accurately calibrated data have been made available and allow the derivation of accurate astrophysical parameters of sources.  Fibers, however, do not ameliorate temporal effects such as differential piston which remain a concern.\\
As most interferometers have been operated in the near-infrared using small telescopes with read-out noise limited detectors, most observations have been carried out in bands a few hundred nanometers wide to ensure sufficient signal. The current generation interferometers like VLTI or Keck benefit from large pupils but will still observe in wide bands at maximum sensitivity. It is the purpose of this paper to study the sensitivity of wide band $V^2$ estimators to parameters such as spectral resolution or instrument defects. Because piston is achromatic, the study of the sensitivity to turbulent optical path difference fluctuations applies equally well to the narrow and wide band cases. These biases must be taken into account when  observations require maximum accuracy as is the case, for example, for direct exoplanet detection. Estimator bias must be taken into account in setting acquisition parameters such as exposure time or fringe frequency and interferometer set-up such as spectral resolution. \\
The general formalism used in this paper is presented in Section~\ref{sec:formalism}. The effect of spectral bandwidth on $V^2$ estimators is discussed in Section~\ref{sec:resolution}. We investigate the influences of longitudinal chromatic dispersion in Section~\ref{sec:dispersion} and of finite scan length in Section~\ref{sec:scanlength}. We simulate the biases and errors caused by optical path difference turbulent fluctuations in Section~\ref{sec:piston}. 

\begin{figure}[t]
\includegraphics[angle=-90, width=16cm]{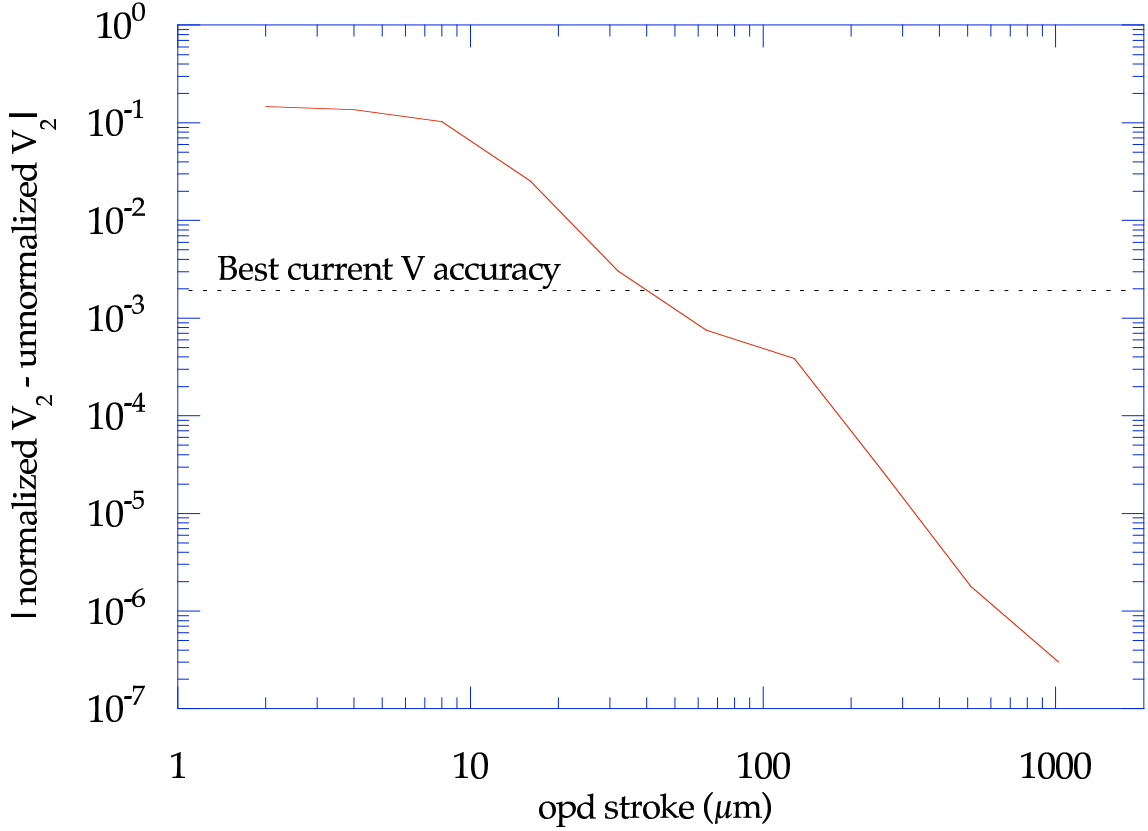}
      \caption{Absolute difference between normalized and unnormalized $V_2$ estimators for a 100\% visibility as a function of opd stroke. Calculation has been performed in the full K band for a G5~III star. Considering the current level of visibility accuracy of single-mode interferometers, the normalization factor can be dropped without causing a significant bias if strokes larger than 40\,$\mu$m are used.} 
         \label{fig:figure_0}
   \end{figure}

\section{General formalism}
\label{sec:formalism}
\subsection{The interferometric equation}
We describe in this section the main equations required to produce the results presented in the next two sections. We have chosen the formalism of co-axial interferometers in which the optical path difference modulation is produced by moving optics (e.g. the Michelson interferometer) as opposed to multi-axial interferometers in which the optical path difference varies with spatial coordinates in the image. The nature of the effects discussed in this paper is generally the same for the two types of interferometers. Multi-axial interferometers are however more sensitive to optical path difference fluctuations and to bandwidth effects. The interferogram being multiplied by a non-flat window (the diffraction pattern for a two-aperture Young's experiment), fringe modulation degrades for non zero optical path differences when wide bandwidths are used. It is not the scope of this paper to discuss this issue and we will focus on co-axial interferometers for which the acquisition window is flat (the interferogram is multiplied by a temporal window of unit height). We choose a beamsplitter-type combiner with two complementary outputs. \\
We note $S(\sigma)$ the spectrum of the source as a function of wavenumber $\sigma$ and $F(\sigma)$ the transmission of the detector filter or the spectrometer transmission for a given channel. Assuming perfectly balanced beams and perfectly corrected turbulence over each pupil, the intensity of one of the two outputs can be written:
\begin{eqnarray}
\label{eq:interferogramme}
I(x=v.t)=& & \int_0^{+\infty}S(\sigma)F(\sigma)\,d\sigma  \\ \nonumber
              &+&\int_0^{+\infty}S(\sigma)F(\sigma)V(\sigma) \\ \nonumber
              &  & \times \cos\left(2\pi\sigma x+\phi(\sigma)+2\pi\sigma p(x)\right)\,d\sigma
\end{eqnarray}
where $x$ is the optical path difference (opd) between the two beams, $t$ is time, $v$ is the velocity of the optical path difference or fringe speed (twice the speed of a moving retro-reflector in a single-pass delay line system), $V(\sigma)$ is the visibility modulus as a function of wavelength, $\phi(\sigma)$ is the phase of the visibility and $p(x)$ is the differential piston. The first integral is the total power collected by the interferometer. The second integral is the fringe pattern. Because of the finite bandwidth $\delta\!\sigma$ of the filtering function $F$, the characteristic number of fringes in the interferogram derived from the coherence length equals $\sigma/\delta\!\sigma$. In the near-infrared with filter bands similar in width to the J, H or K photometric bands, the typical number of fringes is smaller than 10. 

In a perfect instrument, the total power can be accurately measured and the interferogram can be normalized to provide a direct measurement of the real part of the mutual complex spatial degree of coherence $\gamma_{12}(x)$:
\begin{equation}
\gamma_{12}^{\mathrm{real}}(x)=\int_0^{+\infty}W(\sigma)V(\sigma)\cos\left(2\pi\sigma x+\phi(\sigma)+2\pi\sigma p(x)\right)\,d\sigma
\end{equation}
where the weighting function $W$ is:
\begin{equation}
W(\sigma)=\frac{S(\sigma)F(\sigma)}{\int_0^{+\infty}S(\sigma)F(\sigma)\,d\sigma}
\end{equation}

In the case of quasimonochromatic light (\cite{goodman1985}), $\gamma_{12}^{\mathrm{real}}(x)$ is the real part of the product of the complex degree of coherence $\gamma (x)$ -- which measures the temporal coherence of the source -- by the complex coherence factor $\mu_{12}$ which contains the spatial information on the source.
From here on, the phase of visibilities is assumed to be zero (or $\pi$ when the sign of the visibility function changes) and $\gamma$ relates to the visibility through:
\begin{eqnarray}
\gamma_{12}^{\mathrm{real}}(x)& = &\int_0^{+\infty}W(\sigma)V(\sigma)\cos\left(2\pi\sigma x+2\pi\sigma p(x)\right)\,d\sigma \\ \nonumber
& = &\int_{-\infty}^{+\infty}V_{\mathrm{e}}(\sigma)\exp\left(-2i\pi\sigma \left(x+p(x)\right)\right)\,d\sigma \\ \nonumber
\end{eqnarray}
where we call $V_{\mathrm{e}}(\sigma)=\frac{1}{2}\left(W(-\sigma)V(-\sigma)+W(\sigma)V(\sigma)\right)$ the extended visibility.
When the piston term is equal to zero, the mutual degree of coherence $\gamma$ is therefore the Fourier transform of the extended visibility. Conversely, the Fourier transform of $\gamma_{12}^{\mathrm{real}}(x)$ is the weighted visibility function versus wavelength, the principle of double Fourier interferometry (\cite{mariotti1988}).


\begin{figure*}[htbp]
\hbox{
\includegraphics[angle=-90, width=8cm]{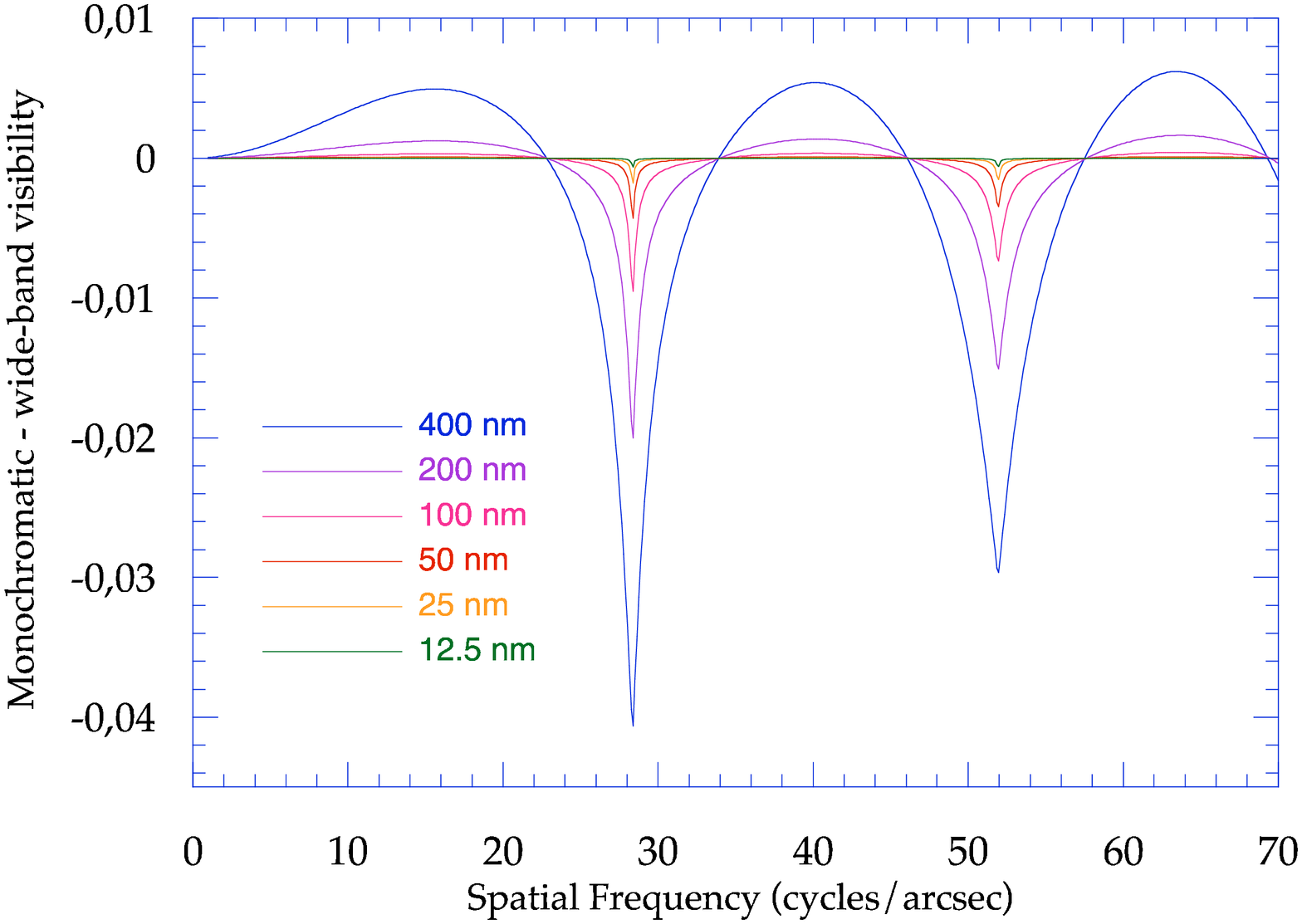}
\hspace{0.1cm}
\includegraphics[angle=-90, width=8cm]{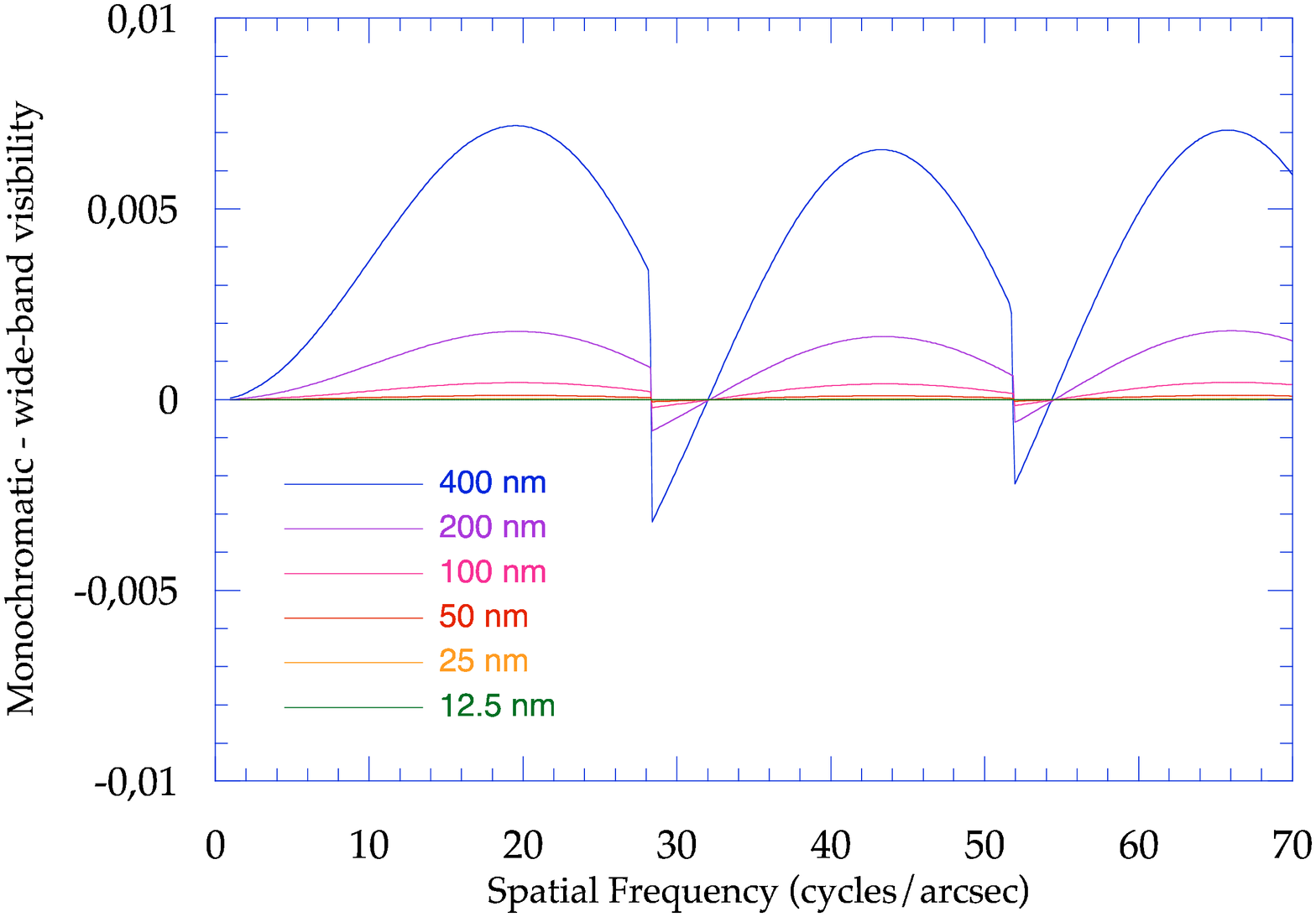}
}
  \caption{Difference between the monochromatic visibility modulus of a 43\,mas uniform disk and the wide-band estimators as a function of spectral bandwidth at 2.2~$\mu$m. $V_1$ is left, $V_2$ is right. The peak errors in  $V_1$ occur at spatial frequencies corresponding to the visibility zeroes for the central wavelength of each filter bandpass.} 
         \label{fig:figure_1}
   \end{figure*}

\subsection{Visibility estimator}
\subsubsection{Continuous signals}
In practice, the phase of interferograms is not accurately known hence it is not possible to measure the complex visibility. The observational goal is to estimate the modulus $V$ of the visibility. Let us first consider the piston-free case. In this special case, the interferometer is a generalized Fourier transform spectrometer working at a non zero baseline. It is possible to measure the weighted visibility modulus as a function of wavelength -- the knowledge of the source spectrum then yields the visibility modulus as a function of wavelength:
\begin{equation}
V(\sigma)=\frac{2}{W(\sigma)} \left| \int_{-\infty}^{+\infty}\gamma_{12}^{\mathrm{real}}(x)\exp\left( 2i\pi\sigma x \right)\,dx \right|
\end{equation}

In presence of additive noise of variance $\Sigma^2$ an unbiased estimator is $\mathcal{V}^2(\sigma)=V^2(\sigma)-\Sigma^2$. This is why we only consider quadratic estimators in the following. 


\subsubsection{Sampled signals}
Real instruments measure samples with spacing $\delta\! x$ on a length $\Delta\! x = N \delta\! x$  yielding the sampled mutual degree of coherence:
\begin{equation}
\gamma_s (x)=\left[ \left[ \gamma_{12}^{\mathrm{real}}(x)\times \shah \left( \frac{x}{\delta\! x} \right) \right] \times \Pi \left( \frac{x}{\Delta\! x} \right) \right] \star \shah \left( \frac{x}{\Delta\! x} \right)
\end{equation}

$\gamma_{12}^{\mathrm{real}}(x)$ is the continuous real part of the complex mutual degree of coherence. The multiplication by the $\shah  \left( \frac{x}{\delta\! x} \right) $ function expresses that it is sampled with a $\delta\! x$ spacing between samples. The finite scan length is mathematically expressed by the multiplication by the $\Pi \left( \frac{x}{\Delta\! x} \right)$ function. This assumes that the time constant of the detector is negligible compared to the reciprocal of the sampling frequency.

 A discrete spectrum is obtained thanks to the convolution by the $\shah \left( \frac{x}{\Delta\! x} \right)$ function which periodically replicates the sampled temporal signal whose acquisition length $\Delta\! x$ is finite. Replication is necessary to ensure that both $\gamma_s$ and its spectrum $V_s$ are discrete. In addition, $V_s$ is also periodic. This is the classical mathematical expression of discrete Fourier transforms of sampled physical signals. 

For integrating detectors (most if not all infrared detectors) an extra convolution of $\gamma_{12}^{\mathrm{real}}(x)$ by the  door function $\Pi \left( \frac{x}{\delta\! x} \right)$ is required:

\begin{equation}
\gamma_s (x)=\left[ \left[ \left[ \gamma_{12}^{\mathrm{real}}(x)  \star \Pi \left( \frac{x}{\delta\! x} \right) \right] \times \shah \left( \frac{x}{\delta\! x} \right) \right] \times \Pi \left( \frac{x}{\Delta\! x} \right) \right] \star \shah \left( \frac{x}{\Delta\! x} \right)
\end{equation}

We only consider the case for which sampling and integration times are equal. In visibility space, the spectrum is filtered by the sinc function $\widetilde{ \Pi}(\sigma \delta\! x)$ whose minimum value is $2/\pi$ at Nyquist frequency. The impact of integration on the visibility estimator is to reduce the modulation and therefore to require a normalization by an ad-hoc factor. This factor depends upon the source spectrum and visibility distribution but also on the sampling frequency. The larger the sampling frequency with respect to the fringe frequency the smaller the contrast reduction and therefore the sensitivity of the estimator to spectral features. When fringe energy is spread over a large bandwidth  because of vibrations or atmospheric piston, the filtering will remove energy and therefore reduce contrast. This effect appears when the source of bias is piston and is therefore considered in Section~\ref{sec:piston}. Integration is otherwise not taken into account in the following for the sake of simplicity.

Assuming that signals are sampled but not integrated, $V_s$ is equal to: 

\begin{equation}
V_s(\sigma)=\left[ \left[ V_e(\sigma) \star \widetilde{ \Pi}(\sigma \Delta\! x) \right] \times \shah(\sigma \Delta\! x) \right] \star \shah(\sigma \delta\! x)
\end{equation}

where $\widetilde{ \Pi}$, the Fourier transform of $ \Pi$, is a $sinc$ function of width $1/\Delta\! x$.
The resolution of the measured visibility spectrum is therefore directly proportional to the interferogram length, a well-known fact in Fourier transform spectroscopy: $R=\sigma . \Delta\! x$. In the following, the spectral information being replicated, we only consider the first $N$ samples :
\begin{equation}
\label{eq:sampled_V}
V_s(k \delta\!\sigma)= \left[ V_e(\sigma) \star  \widetilde{ \Pi}(\sigma \Delta\! x) \right] (k \delta\!\sigma ),\;\;\; k=0,\dots,N-1
\end{equation}
with $\delta\!\sigma=\frac{1}{\Delta\! x}$.

\section{Effect of spectral bandwidth}
\label{sec:resolution}
In this section we assume an infinite scan length and study the effect of spectral bandwidth which in practice cannot be negligible. The assumption of infinite scan length is equivalent to assuming that the source spectrum only contains spectral features whose width is larger than the Fourier spectral resolution $\frac{\Delta\! x}{\lambda}$. Cases for which spectral features are smaller or close to the spectral resolution need to be modeled if  high accuracy is required. \\
This latter case is also a source of bias as illustrated by the  following example. Suppose a resolved source of continuum with a 50\% flat visibility across the bandpass, and an unresolved spot in a 10 km/s line with the source and the spot emitting equal fluxes. The unresolved spot has a visibility of 100\%. The line gets averaged with the continuum in the convolution by the spectral response of the interferometer. Assuming a scan length of 100~$\mu$m at 2~$\mu$m wavelength, hence a resolution of 50, the width of the line is 1/600 of a spectral element. The visibility being the linear average across a spectral element, it is of 0.50042 instead of 0.5 with continuum emission only hence a bias of 0.08\%. \\
In the following we assume a smooth visibility spectrum and investigate the impact of bandwidth on estimators. 
\\

 With this assumption, Equation~\ref{eq:sampled_V} has the simplified expression:
\begin{equation}
V_s(k \delta\!\sigma)=V_e (k \delta\!\sigma),\;\;\; k=0,\dots,N-1
\end{equation}
We investigate the influence of the spectral bandwidth on the wide-band visibility estimators. Noise is neglected in the following but we use squared moduli to remain in the framework of an unbiased estimator in presence of additive noise.\\
Two visibility modulus estimators can be built depending on whether real or imaginary parts of the extended visibility are summed prior to being squared:

\begin{eqnarray}
\label{eq:estimateurs}
V_1^2 & = & \frac{4}{\sum_{k=0}^{N-1} W^2(k \delta\!\sigma) \delta\!\sigma}\sum_{k=0}^{N-1}|V_e (k \delta\!\sigma)|^2 \delta\!\sigma  \\ \nonumber
V_2^2 & = &  \frac{4}{\left[\sum_{k=0}^{N-1} W(k \delta\!\sigma) \delta\!\sigma\right]^2} \times  \\ 
& & \;\;\;\;\;\;\;\; \left[ \left[ \sum_{k=0}^{N-1}\Re(V_e (k \delta\!\sigma)) \delta\!\sigma\right]^2+  \left[ \sum_{k=0}^{N-1}\Im(V_e (k \delta\!\sigma)) \delta\!\sigma\right]^2 \right] \nonumber
\end{eqnarray}

Both estimators have drawbacks and advantages. The first one, $V_1^2$, is insensitive to atmospheric or instrumental differential phase effects such as chromatic phase dispersion generated by air at visible wavelengths or by unmatched dispersion in single-mode fibers - it is for these reasons that it has been so widely used. Yet, it has to be normalized by the shape factor $\sum_{k=0}^{N-1} W^2(k \delta\!\sigma) \delta\!\sigma$  which depends on the source spectrum (\cite{foresto1997}), hence possible bias effects must be considered. The second estimator, $V_2^2$, is adapted from the multi-speckle mode visibility estimator of \cite{berio1999}. It is sensitive to differential phase and requires these effects to be negligible or at least stable enough to be calibrated. As $V_2^2$ applies a summation over linear quantities, the normalization factor $\left[\sum_{k=0}^{N-1} W(k \delta\!\sigma) \delta\!\sigma\right]^2$  is very close to 1 if long opd stroke scans are recorded. This makes this estimator much less sensitive to spectral features than $V_1$. For example, the bias introduced by deep absorption bands is negligible in the case of $V_2$ compared to $V_1$. The normalization factor cannot however be neglected if short scan lengths are used as shown on Fig.~\ref{fig:figure_0}.   We have plotted the difference between $V_2$ computed with and $V_2$ computed without the normalization factor as a function of opd stroke. 
 We have used the spectrum of a G5~III star in the K band.
Considering the current level of visibility accuracy of single-mode interferometers of $0.2$~\% (\cite{perrin2004}), the factor can be dropped without causing a significant bias if strokes larger than 40\,$\mu$m  are used in this case. The required opd stroke varies with the source spectrum. It is 60\,$\mu$m for a flat spectrum which is the most extreme example.
When the filtered spectra of the calibrator and science target are similar, the normalization factors disappear in the interferometric calibration. Yet, the ratio of the normalization factors should be estimated when it is a potential critical source of bias. In the case of integrating detectors, the contrast loss due to the averaging of the fringe modulation on a finite time scale is directly taken into account by the normalization factor. For monochromatic signals it is directly equal to $\left( \frac{\pi \sigma \delta\!x}{\sin (\pi \sigma \delta\!x)} \right)^2$ with the particular value of $\pi^2/8$ at four samples per fringe. The signal-to-noise ratios of both estimators are calculated in the Appendix. $V_1$ is more sensitive than $V_2$ by a factor 2 to 3 under realistic conditions. \\

For both $V_1$ and $V_2$, the estimated visibility modulus is not exactly the object visibility modulus measured at the effective wavelength of the instrument. 
 The reason is that the averaged squared visibility function in a band of finite bandwidth is different from the visibility function. This is particularly true when the visibility function flips sign across the band.
This effect has already been described in recently published papers for estimator $V_1$ (\cite{kervella2004,perrin2004}). We have illustrated this with the example of a 43\,mas uniform disk observed at a wavelength of 2.2\,$\mu$m. We have plotted on Fig.~\ref{fig:figure_1} the difference between the monochromatic visibility modulus at the effective wavelength:
\begin{equation}
V(S)=\left| \frac{2J_1(\pi\O S)}{\pi \O S} \right|
\end{equation}
with $S$ the spatial frequency and $\O$ the uniform disk diameter, and the wide-band visibility modulus estimated with the $V_1$and $V_2$ estimators. Bandwidths range from 12.5\,nm to 400\,nm, the full K band. We have used flat spectrum and transmission. The difference between the two estimators is evident. $V_2$ is very close to the monochromatic function even with the largest bandwidth. The difference is at most $0.6\%$ and reaches maxima at the top of visibility lobes or at the inflexion point. The $V_1$ estimator has the same features but the difference is much larger at the location of the nulls of the visibility function with a maximum difference of $4\%$. In both cases, the effect of bandwidth cannot be neglected for bands a few hundred nanometers wide.

\begin{figure}[htbp]
\includegraphics[angle=-90, width=16cm]{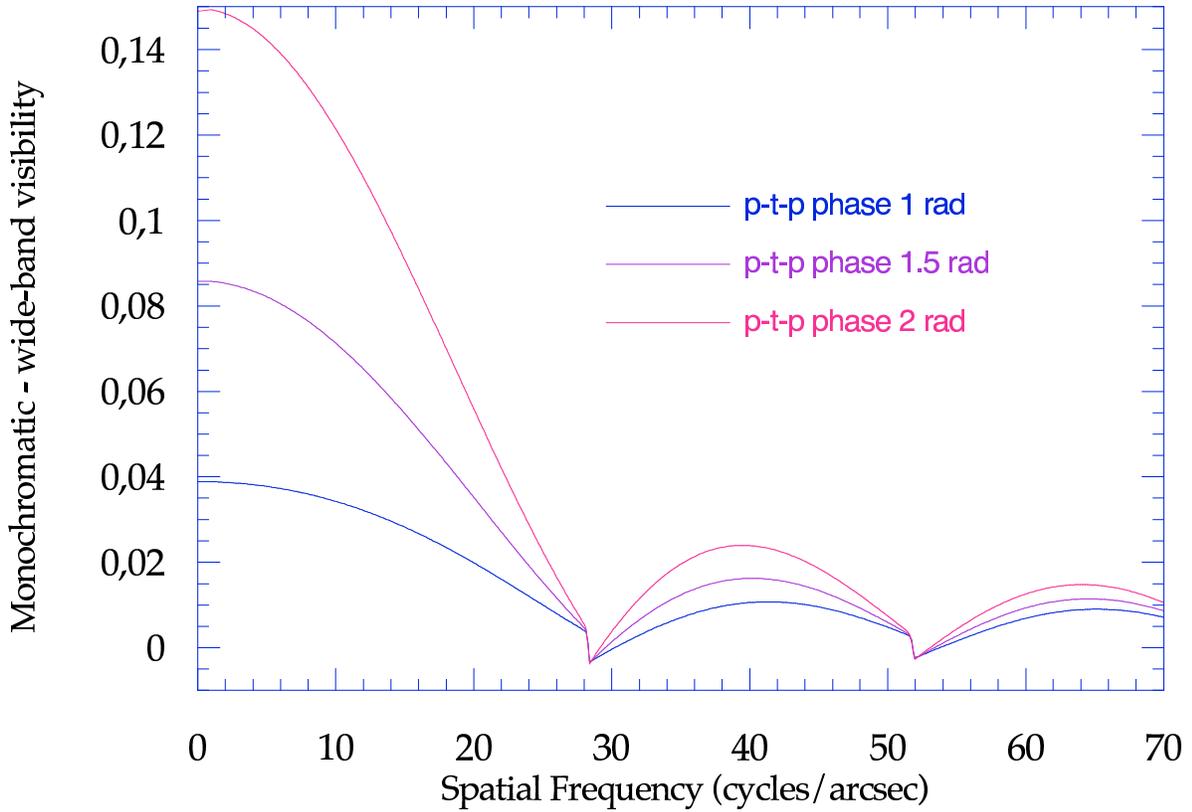}
      \caption{Difference between the monochromatic visibility modulus of a 43\,mas uniform disk and the $V_2$ wide-band estimator as a function of dispersion strength in a  400\,nm wide band around 2.2\,$\mu$m. The dispersion strength is expressed as peak-to-peak phase.}
         \label{fig:figure_2}
   \end{figure}

\begin{figure*}[htbp]
\hbox{
\includegraphics[angle=-90, width=8cm]{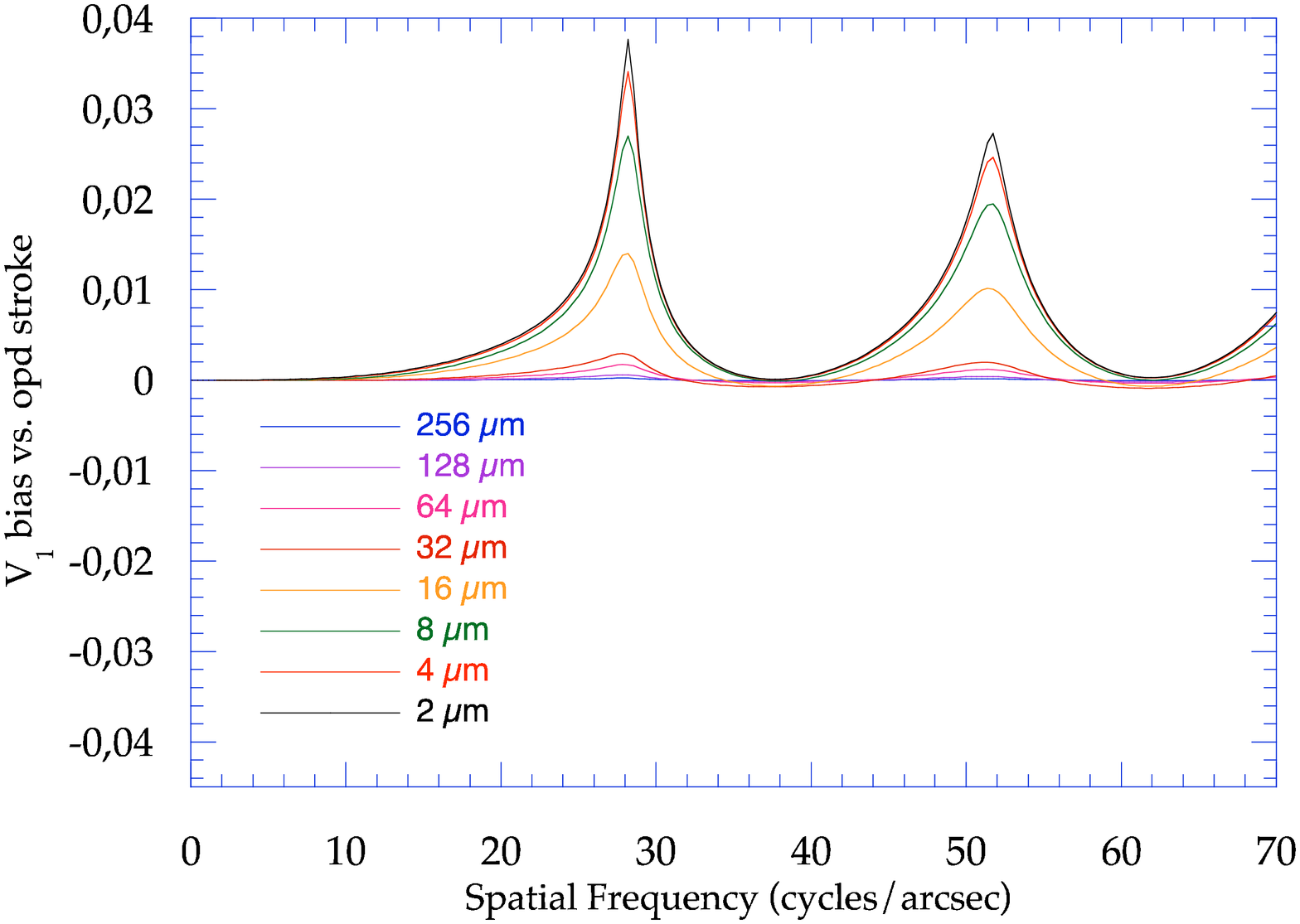}
\hspace{0.1cm}
\includegraphics[angle=-90, width=8cm]{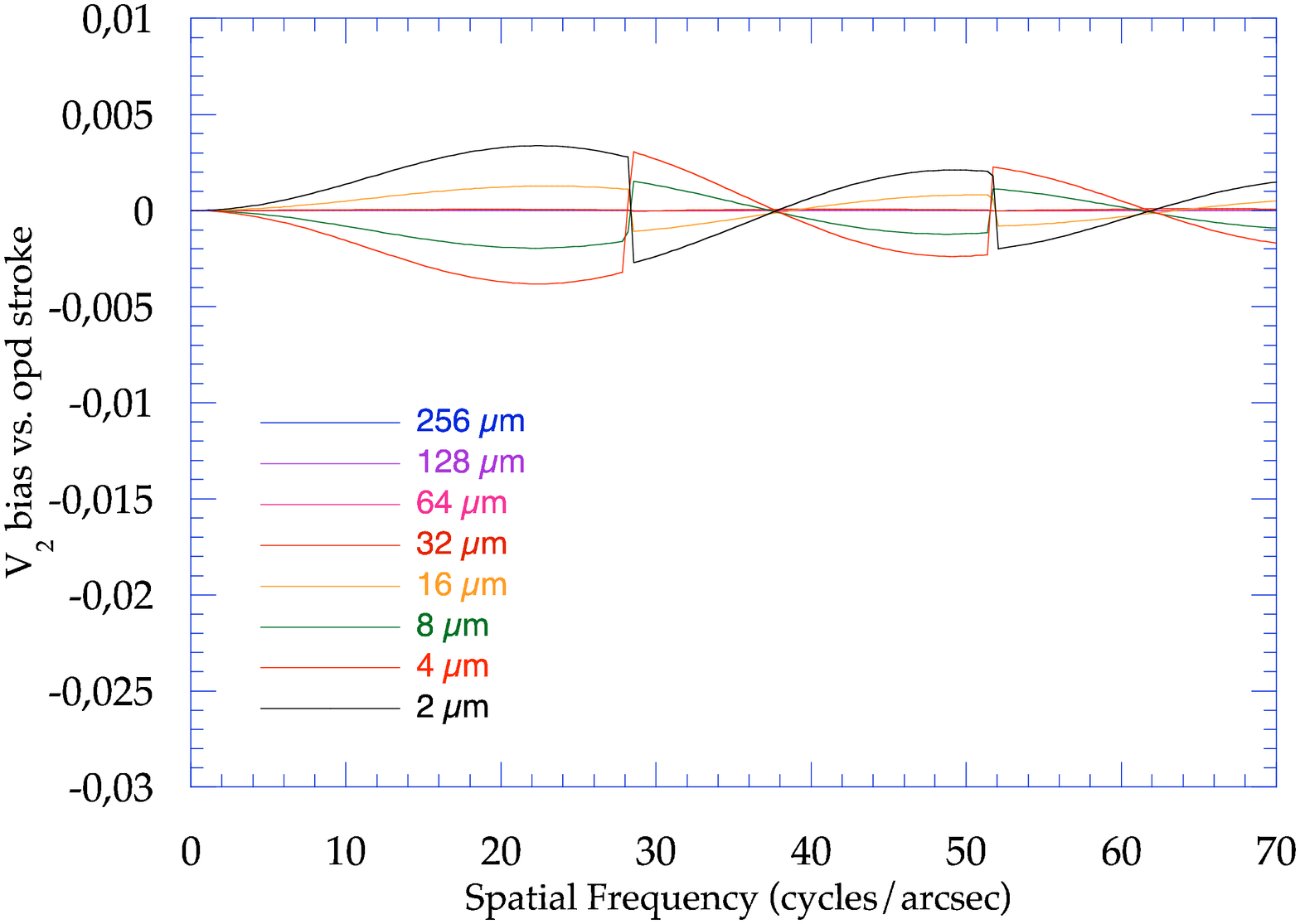}
}
 \caption{Bias of visibility estimators with respect to an infinitely long scan as a function of opd stroke. The source is a 43\,mas uniform disk observed in a 400\,nm wide band centered on 2.2\,$\mu$m. $V_1$ is left, $V_2$ is right. }
         \label{fig:figure_3}
   \end{figure*}


\section{Effect of chromatic dispersion}
\label{sec:dispersion}
When two interferometric beams are crossing different thicknesses of glass or air (at the blue end of the spectrum) longitudinal chromatic dispersion (or group delay dispersion) is produced. An additional phase term needs to be added to the fringe spectrum which takes the simple form:
\begin{equation}
\phi(\sigma)=\frac{1}{2}D\left( \sigma-\sigma_0 \right)^2
\end{equation}
where $\sigma_0$ is the center of the band and where the 0$^{\mathrm {th}}$ and 1$^{\mathrm {st}}$ phase orders are set to 0 (the interferogram is centered). $D$ is the dispersion coefficient (the second derivative of the phase) and can be expressed in units of $\mu$rad\,cm$^2$. The reader is referred to \cite{foresto1995} for more details on the notations and for a study of single-mode fiber dispersion for astronomical interferometry. The differential phase produced by dispersion has no effect on the $V_1$ estimator by definition when scan length is infinite or when spectral resolution is large enough that phase is constant across a spectral element. 
In wide-band, the $V_2$ estimator gets biased by dispersion as is shown in Fig.~\ref{fig:figure_2} (we took a 400\,nm bandwidth centered on 2.2\,$\mu$m). We have chosen to represent cases for which the peak-to-peak phase produced by dispersion accross the spectrum is 1, 1.5 and 2~radians. Such cases can be encountered at short wavelengths because of the dispersion of the refractive index of air or when using fibers if arms are at unequal temperatures. A 1\,K temperature difference over 1\,km of fluoride fiber will produce a 10\,mm difference of glass optical path length, with a 1.2~radian peak-to-peak phase at 2.2~$\mu$m (\cite{kotani2004}). Any uncontrolled temperature variation will introduce visibility biases with the $V_2$ estimator. Or equivalently at shorter wavelengths, different air pathlengths will cause different dispersions which require a calibration for $V_2$. A calibration can be achieved by measuring the $V_2$ response on a point source or alternatively by measuring the dispersion law and correcting for it in the $V_2$ estimator. However, the calibration is subject to changes hence to uncalibrated biases. 

\begin{figure*}[t]
\hbox{
   \includegraphics[angle=-90, width=8cm]{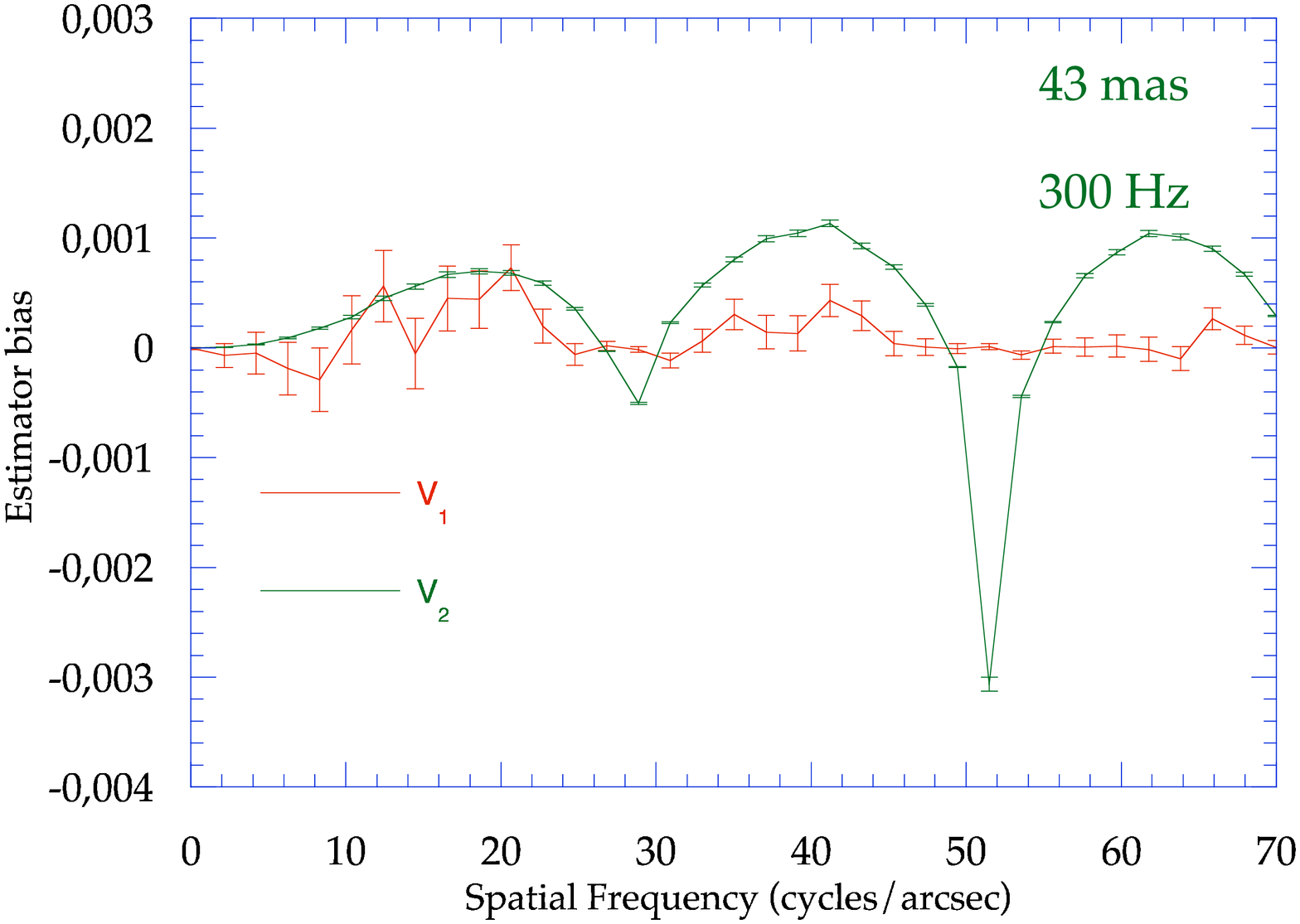}
   \hspace{0.1cm}
   \includegraphics[angle=-90, width=8cm]{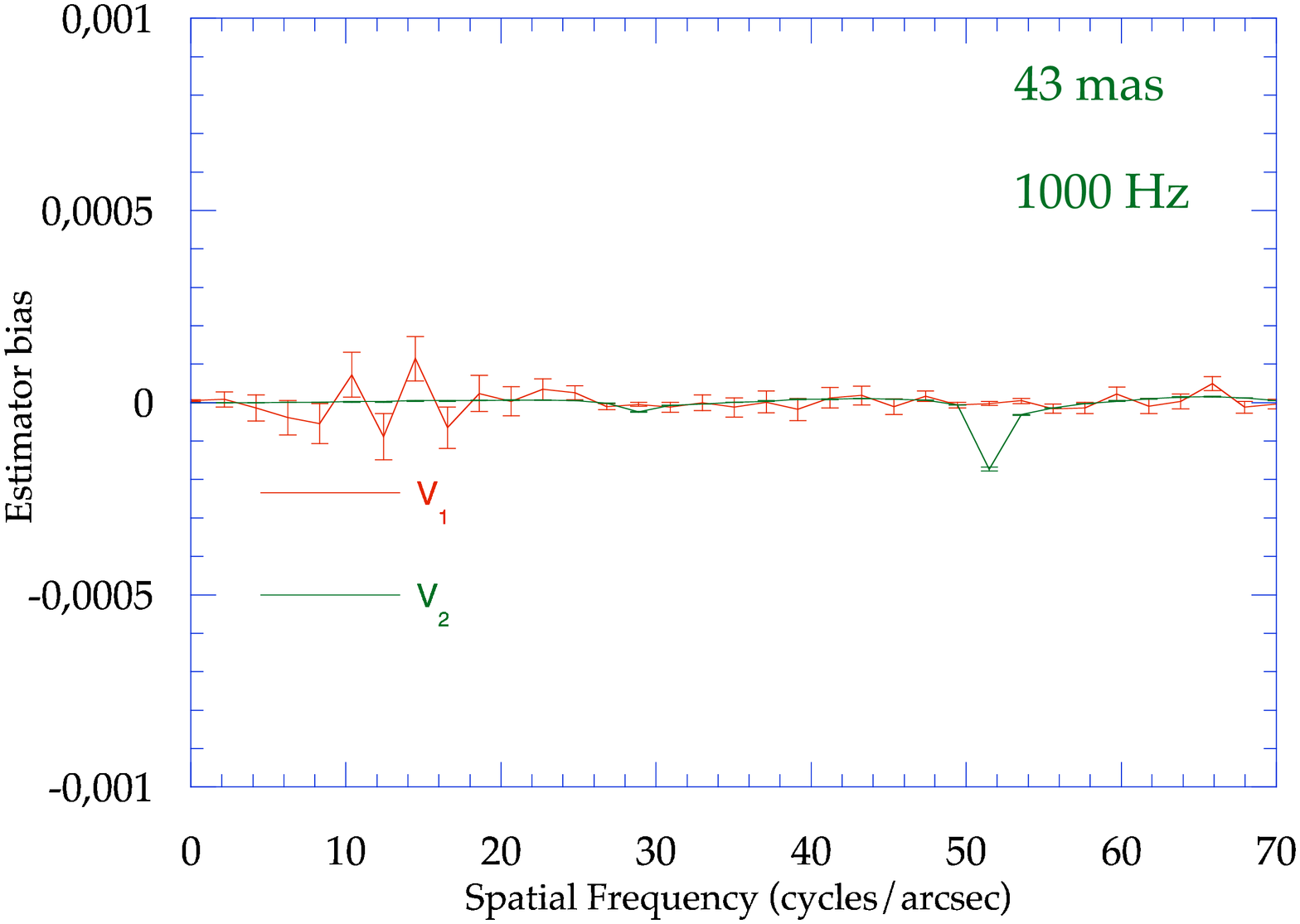}
}
\vspace{0.4cm}
\hbox{
 \includegraphics[angle=-90, width=8cm]{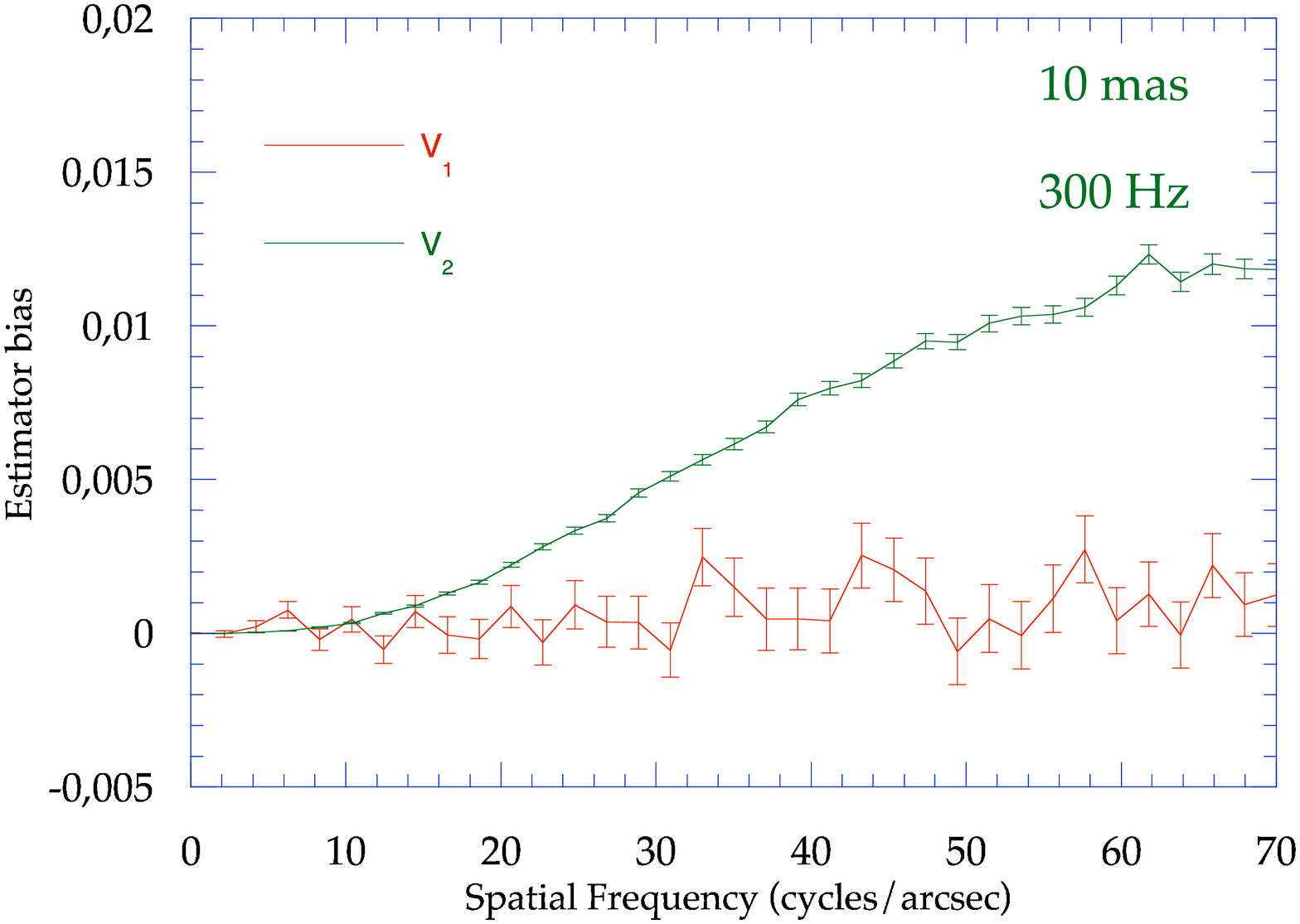}
   \hspace{0.1cm}
   \includegraphics[angle=-90, width=8cm]{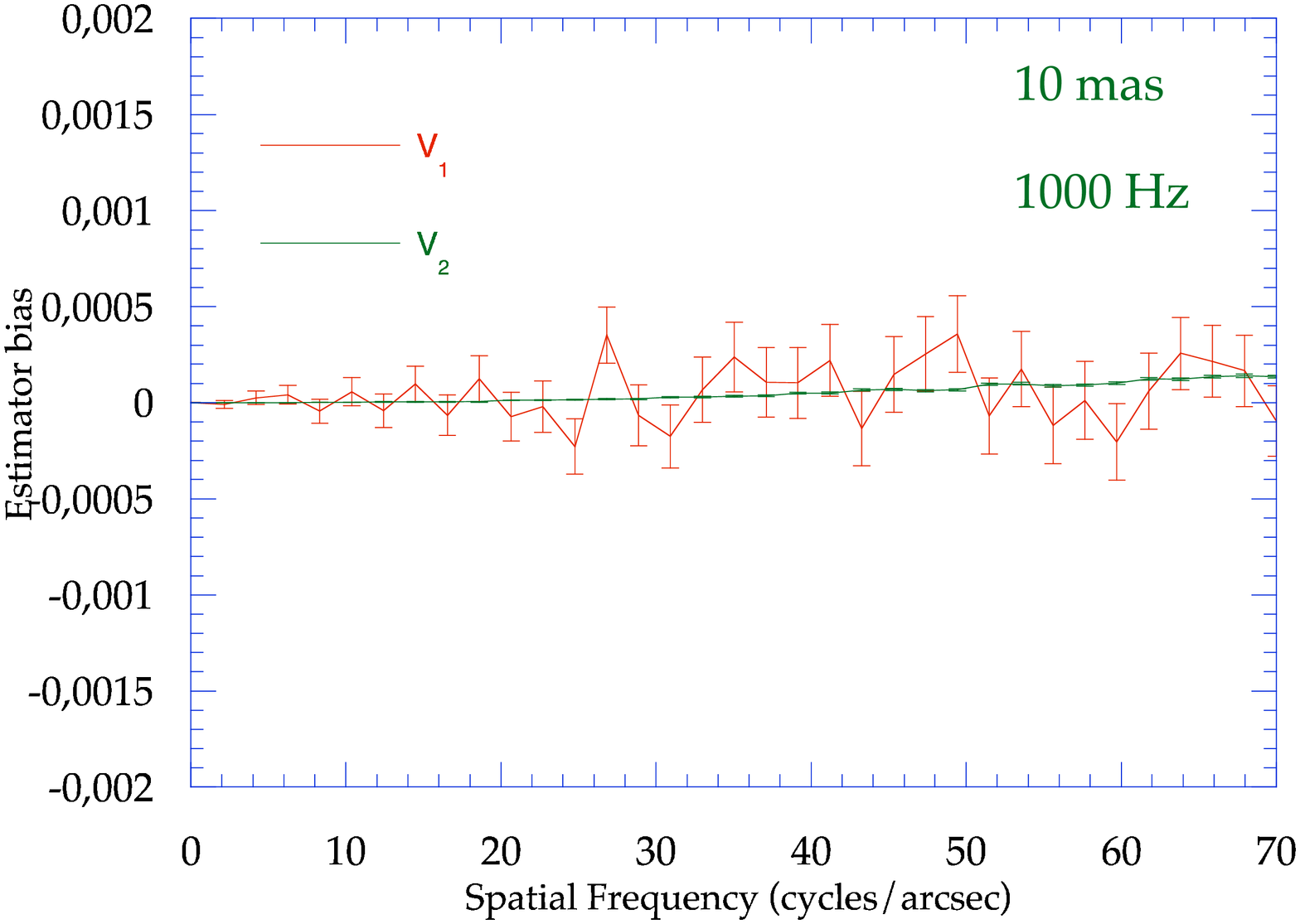}
}
   \caption{Simulation of the noise and bias produced by piston on a 43 and a 10\,mas source observed in a 400\,nm wide band centered on 2.2\,$\mu$m. Two fringe frequencies have been simulated: 300 and 1000\,Hz. See text for more details on piston sequences simulation.}
   \label{fig:figure_4}
\end{figure*}

\section{Effect of finite scan length}
\label{sec:scanlength}
With temporal coding, data processing is very similar to that of Fourier transform spectrometers -- after Fourier transforming the interferograms, the frequency samples  have a resolution proportional to the scan length. 
As a consequence, the measured extended visibility spectrum is the convolution of the real extended visibility by the Fourier transform spectrometer response as expressed in Equation~(\ref{eq:sampled_V}). 
Because of the convolution, measured Fourier components are the weighted sums of spectral elements. The $V^2_1$, for example, is therefore not the pure quadratic sum of spectral elements. 
A bias is therefore to be expected at spatial frequencies where the visibility varies quickly or flips sign. This is illustrated in Fig.~\ref{fig:figure_3} where the biases of the visibility estimators for a 43\,mas uniform disk observed in a 400\,nm wide band centered on 2.2\,$\mu$m  with respect to infinitely long scan estimators
have been plotted for different opd stroke values ranging from 256 to 2\,$\mu$m.
When the stroke is 2\,$\mu$m the estimator at 2\,$\mu$m wavelength is the famous ABCD estimator. As expected, the $V_1$ estimator bias is large where the visibility function of a 43\,mas uniform disk comes to a null. The bias becomes lower to the best current accuracy of 0.2\% for scan lengths longer than 64\,$\mu$m. The $V_2$ estimator naturally turns out less sensitive to the effect as the spectral components are first linearly co-added before squares of the imaginary and real parts are summed. Except for very short scans, the bias is always negligible. This calculation shows however that the estimator needs to be properly designed to not introduce a bias in the interpretation of visibilities if short scan  lengths are preferred to longer ones. This particularly holds for the multi-axial combination for which a few fringes across the diffraction pattern are measured.


\section{Effect of differential piston}
\label{sec:piston}
Unlike dispersion effects or purely acquisition or instrument dependent biases of the estimators, errors due to piston cannot be calibrated so easily.  Piston errors have coherence times of a few milliseconds and their characteristics are not stationary. The best solution is to include a fringe tracking system that will correct for the lower frequency part of the piston. Still, with such a system, only piston errors in the band pass of the fringe tracker are suppressed. Higher frequency errors will blur the fringes during exposures and reduce the contrast -- a source of bias if the fringe tracker performance on the source and on the calibrator are different or if turbulence has evolved between measurements. An alternative is to scan fringes fast enough that piston errors can be frozen during an acquisition at the cost of sensitivity, which is equivalent to benefitting from a fringe tracker with a high band pass. We investigate here the influence of piston in this latter case. 

We first have chosen to simulate piston and data acquisition sequences which correspond to real data acquired on the supergiant $\alpha$~Orionis and its calibrators and reported in \cite{perrin2004}. We have therefore considered two diameters: 43 and 10\,mas and two fringe frequencies ($v/\lambda$ with the notations of Section~\ref{sec:formalism}): 300 and 1000\,Hz. Typical seeing conditions are on the order of 1\,arcsec with a wind velocity of 10\,m/s giving $r_0=45$\,cm (i.e. exactely the size of the 45~cm telescopes) and a coherence time for turbulent phase of 45\,ms in the K band. We have simulated scan lengths of 64\,$\mu$m with 4 samples per fringe. For the fringe frequencies of  300 and 1000\,Hz,  99 and 29.7\,$\mu$m of opd are respectively scanned in a coherence time of 45\,ms. The piston is therefore mostly frozen with the higher frequency and only partially with the smaller one. However, in the latter case, the whole 64\,$\mu$m acquisition length is required to avoid biases due to finite scan length. Piston stability is primarily required over the coherence length which contains most of the fringe power. No additive noise has been injected in the simulations -- only pure piston noise is considered here. Visibilities have been measured for each simulated interferogram using the $V_1$ and $V_2$ estimators. 1000 interferograms are processed for each spatial frequency sample. The piston sequences are computed using the method of \cite{perrin1997}. The baseline used to compute the piston rms and power spectral density varies with spatial frequency according to: $B=\lambda . S$. The lower the spatial frequency the smaller the piston rms. This is a little pessimistic for the longest baselines as recent studies have shown that the outer scale of turbulence at current best sites is about 20~m (see e.g. \cite{martin2000}) inducing a saturation of the rms piston to less than 20~$\mu$m (\cite{davis1995}) or equivalently at baselines of order 20~m depending on models. In our simulation the piston rms does not saturate and linearly increases with baseline up to 35~m. Note that the rms of piston is 0 at 0 spatial frequency.\\

The result of the simulation is presented in Fig.~\ref{fig:figure_4}. The differences between unpistoned and pistoned estimated visibilities are plotted. The error bars are the standard deviations computed from the 1000 interferograms. Estimates are biased when differences are more than 3~$\sigma$ from 0. We conclude from the simulation that $V_1$ is far less biased by piston than the $V_2$ estimator. At the level of our simulation, piston can be considered a zero-mean noise for the $V_1$ estimator, biases being smaller than 0.1\% at 300\,Hz fringe frequency. This is intuitively reasonable as the $V_1$ estimator is not sensitive to interferogram shifts and by extension poorly sensitive to the first orders of piston whereas the $V_2$ estimator being a coherent estimator is sensitive to all piston orders including the lowest ones. 

Piston is clearly a bias for $V_2$ at low fringe frequency.  Interestingly, the statistical scatter in the simulation for $V_2$ is always much smaller than for the $V_1$ estimator -- perhaps thanks to the coherent averaging of spectral elements.

At higher fringe frequency however, when piston is almost completely frozen, the bias on $V_2$ is usually far less or less than 0.01\%. In practice, only bright sources in good weather conditions can be observed at high fringe frequency. 
With the parameters used for our simulation, the statistical error on visibilities generated by piston is of 0.1\% or larger if less than 1000 scans are acquired. This is in accord with current limits in visibility accuracies measured by single-mode interferometers which can thus be considered unbiased by piston if suitable fringe frequencies are chosen - as in, for example, \cite{perrin2004}.

\begin{figure*}[t]
\hbox{
   \includegraphics[angle=-90, width=8cm]{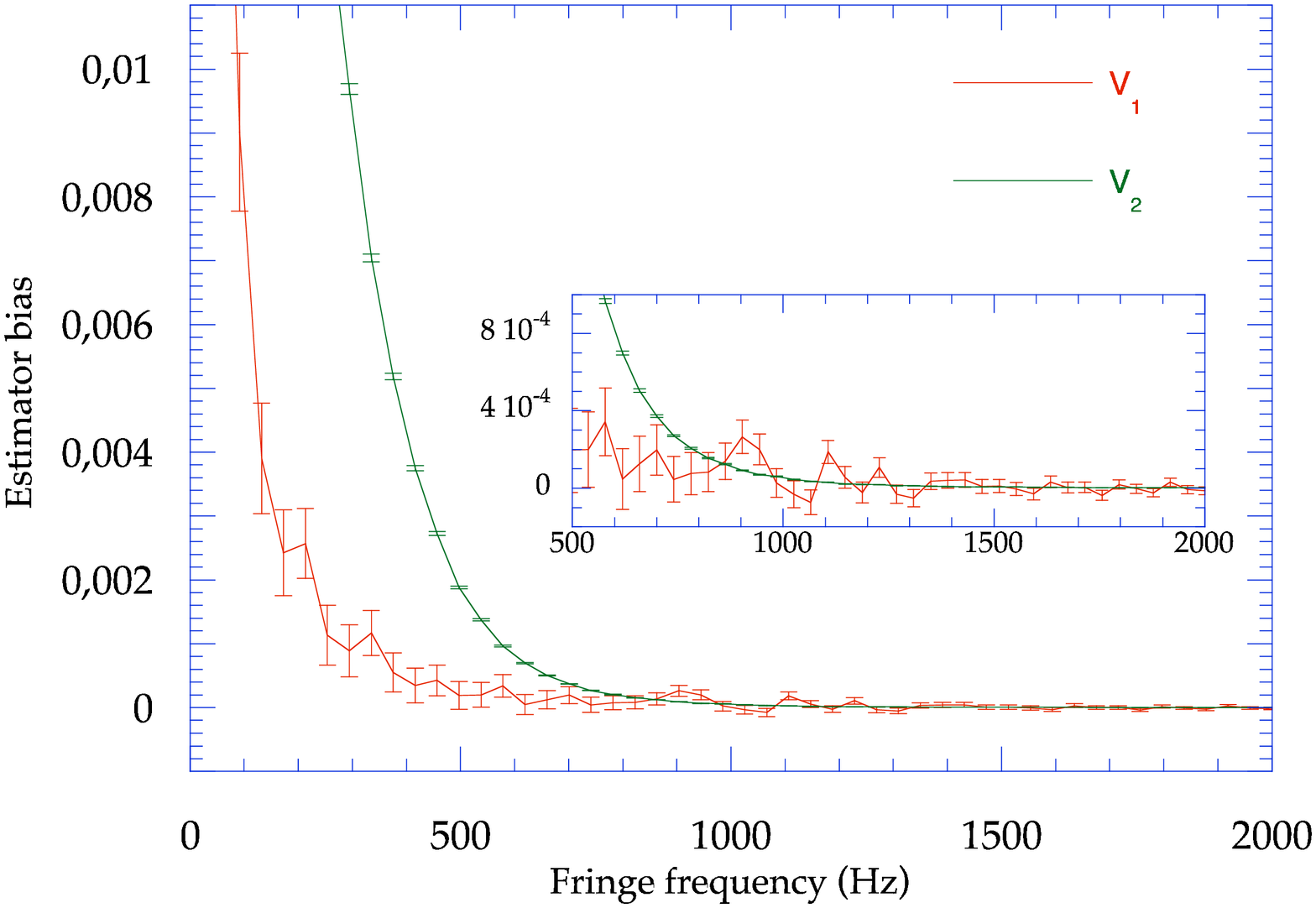}
\hspace{0.1cm}
   \includegraphics[angle=-90, width=8cm]{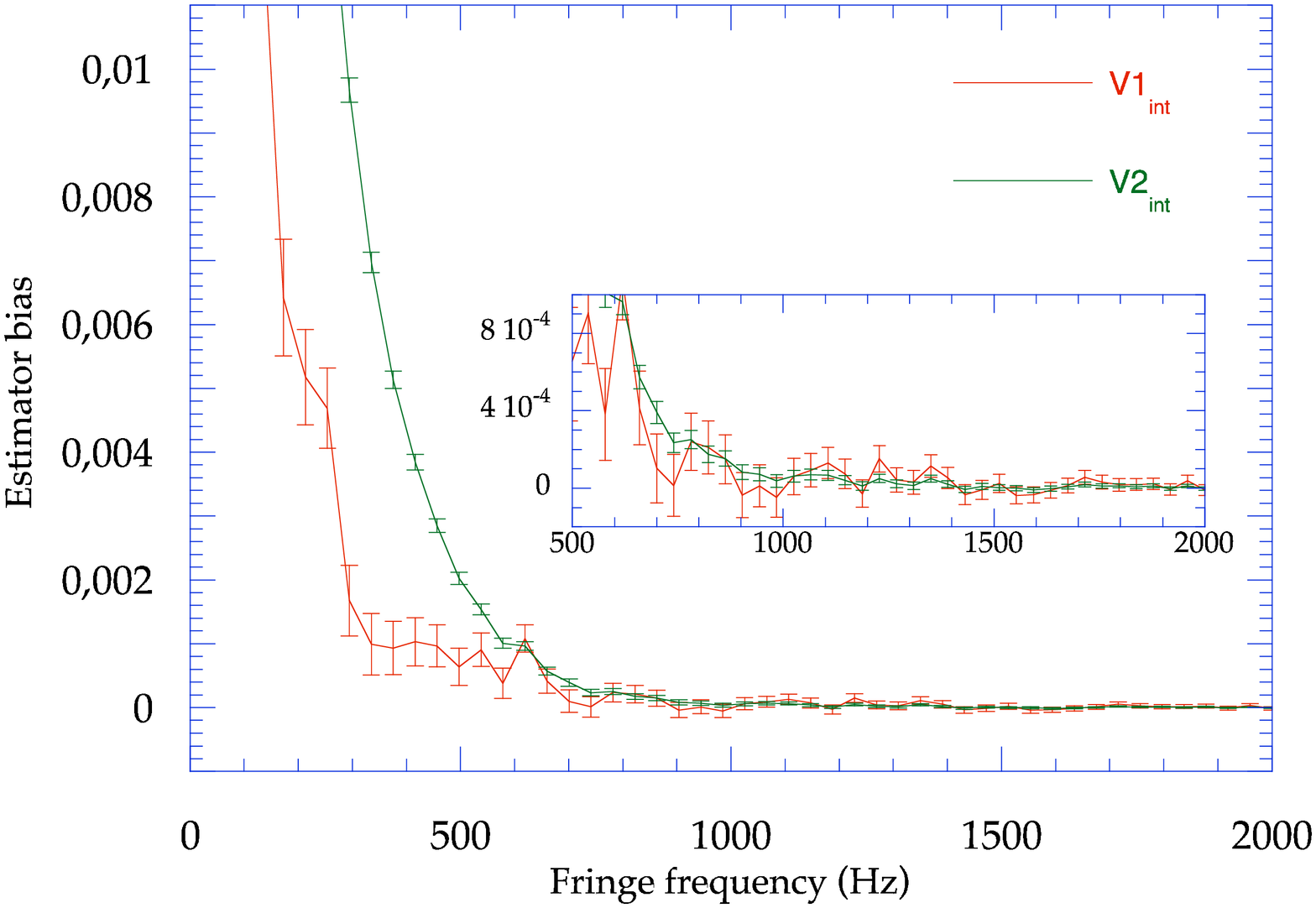}
}
 \caption{Simulation of the noise and bias produced by piston on estimators for a point-like source. See text for parameters. The insets are close-up views of the curves between 500 and 2000~Hz. Integration is taken into account for the right panel whereas signals are just sampled in the left panel. }
   \label{fig:figure_5}
\end{figure*}

It turns out from this discussion that the two estimators have different properties depending on fringe acquisition parameters. We have computed the bias and the statistical error for the two estimators as a function of fringe frequency for the following set of parameters (Fig.~\ref{fig:figure_5}, left panel):  1~arcsec seeing, infinite outer scale, 10~m/s wind velocity, K band, 64~$\mu$m scan length, 4 samples per fringe, 45~cm telescopes, unresolved source (100\% visibility), 40~cycles/arcsec spatial frequency (18~m baseline), 10000 interferograms. The following conclusions do not depend upon bandwidth and also apply to narrower bands. The $V_1$ estimator is biased at more than 1\% below 100~Hz fringe frequency. Same for $V_2$ but below 300~Hz. $V_1$ is unbiased above 300~Hz. Above 800~Hz, $V_2$ becomes much more accurate than $V_1$ and reaches accuracy levels below $10^{-5}$. As a conclusion, $V_1$ is better suited for lower fringe frequencies than $V_2$ but $V_2$ provides much better accuracies at larger fringe frequencies or with a fringe tracker with a band pass larger than 800~Hz.

 The estimators have also been simulated taking integration into account between consecutive samples. The general tendency is the same except at frequencies below 200\,Hz for which the bias is more important. The integration slightly blurs the fringes if the fringe frequency is too low. The effect is more important for $V_1^2$ than for $V_2^2$ which performs a linear sum of Fourier components. Results are displayed in the right panel of Fig.~\ref{fig:figure_5}.

These conclusions can be extended to other parameter values. First, the noise and bias produced by piston are multiplicative. The piston effect can be scaled from Fig.~\ref{fig:figure_5} for any visibility. The fringe frequency axis can be rescaled as a function of wind speed according to: $f\times \left( \frac{10 \mathrm{m/s}}{v_{\mathrm{wind}}}\right)$. We have chosen a baseline of 18~m which is supposed to be just below the saturation of the piston rms. Longer baselines should therefore generate the same amount of piston whereas shorter baselines will have less piston making this simulation a little bit pessimistic. Scaling this graph to better or poorer seeing is unfortunately tricky as the link between visibility errors due to piston and seeing is not linear. In first approximation this graph can be considered for any seeing value around 1~arcsec. The size of telescopes and the baseline will change the cut-off frequencies of the piston spectrum whose consequence is difficult to guess from Fig.~\ref{fig:figure_5}. In the same vein as for seeing, results from this graph should be considered applicable in first approximation. New simulations should be considered for significantly different values for these parameters.

This study shows that the fringe frequency should therefore be tuned with respect to the piston conditions and the accuracy to be achieved in the same way as the number of exposures is set in order to achieve the required signal-to-noise ratio. The visibility estimator should also be chosen accordingly. 


\section{Conclusion}
We have shown that visibility estimators must be computed with care for low spectral resolution measurements. We have compared the performances of two visibility estimators under varied instrumental or turbulent conditions. The $V_1$ estimator tends to be more robust to all effects. It is even intrinsically insensitive to longitudinal dispersion. The $V_2$ estimator can be used if differential piston is frozen and if dispersion effects are either stable or measurable with great accuracy. Besides, the spectral slope of the phase has to be removed accurately. $V_2$ is more sensitive to additive noise than $V_1$ by a factor 2 to 3. A first important conclusion is that the biases of the $V_1$ estimator are smaller than the best accuracies (0.2\%) measured for fringe frequencies larger than 200~Hz.  A second important conclusion is that, although  the $V_2$ estimator is more sensitive to phase defects, it has the potential to reach accuracies as good as $10^{-5}$ for high fringe frequencies or with a fringe tracker with a large bandpass. If the visibility accuracy is limited by piston rather than additive noise then the $V_2$ estimator is better. When the observation is performed with the goal of achieving excellent accuracy, effects should be simulated to tune the acquisition parameters such as the length of the interferograms, the acquisition  frequency and the number of recorded exposures. High dynamic range detections are therefore achievable at this price and with suitably designed instruments. 

\acknowledgments
The authors wish to thank the referee for his valuable comments which were helpful to improve the quality  of the paper.

\appendix 

\section{Estimator signal-to-noise ratio}
The signal-to-noise ratio (SNR) of the ABCD estimator has been derived by \cite{tango1980} for the photon noise case and by \cite{colavita1999} for both photon and detector noise. We present here a derivation of the SNR for the $V_1^2$ and $V_2^2$ estimators for any number of samples. 
\subsection{Sampled interferogram formalism}
Notations are derived from the general interferogram expression of Eq.~(\ref{eq:interferogramme}):
\begin{equation}
I_i=C_i+M_i \;\;\;\; i=0 \dots N-1
\end{equation}
in which $N$ is the number of samples, $C_i$ and $M_i$ are respectively the continuous (unmodulated) and modulated parts of the interferogram in photon units. These quantities are corrupted by both source photon and detector noise - the realization of the latter is noted $b_i$ for each $i$. Noise-free quantities are noted (ensemble average):
\begin{eqnarray}
c_i = \left< C_i \right> \\ \nonumber
m_i = \left< M_i \right>
\end{eqnarray}
In the following, we assume that the subtraction of the continuous part of the interferogram and its normalization to get the expression of estimators as in Eq~(\ref{eq:estimateurs}) are noise-free processes. The justification is that the estimate of the continuous part can be low-pass filtered and contains more photons than the modulated part hence a larger signal-to-noise ratio. 
Let $B_k,  \;\; k=0 \dots N-1$ be the discrete spectrum of the modulated samples:
\begin{equation}
B_k=\sum_{i=0}^{N-1}M_i \e^{-2\imath\pi ik/N}
\end{equation}
The unbiased estimators are proportional to the following expressions (taking into account proportionality factors  does not change the final expression of the SNR):
\begin{eqnarray}
V_1^2 & \propto & \sum_{k=0}^{N-1} \left| B_k \right|^2 -N^2\sigma^2 -N\sum_{i=0}^{N-1} m_i\\ \nonumber
V_2^2 & \propto & \left| \sum_{k=0}^{N-1} \Re(B_k)\right|^2 +  \left| \sum_{k=0}^{N-1} \Im(B_k)\right|^2 -N^2\sigma^2 -N^2 m_0
\end{eqnarray}
In the following, in order to make the reading easier, the $\propto$ signs are replaced by equal signs. The expectations of the estimators are:
\begin{eqnarray}
\left< V_1^2 \right> & = & N\sum_{i=0}^{N-1} m_i^2\\ \nonumber
			      &     & \\ \nonumber
\left< V_2^2 \right> & = & N^2m_0^2
\end{eqnarray}
For the $V_2^2$ estimator not to be biased it is therefore very important that the 0$^{\mathrm{th}}$ sample be centered on the white light fringe or, equivalently, the slope of the spectral phase must be accurately removed.
\subsection{Variance of $V_1^2$}
The bias terms are dropped as they do not contribute to the variance. As a consequence:
\begin{equation}
\var(V_1^2)=\sum_{k=0}^{N-1} \var(\left| B_k \right|^2) + \sum_{k\neq q; k,q =0}^{N-1}\covar(\left| B_k \right|^2, \left| B_q \right|^2)
\end{equation}

which can be rewritten as:
\begin{equation}
\var(V_1^2)=\sum_{k,q}\covar(\left| B_k \right|^2, \left| B_q \right|^2)
\end{equation}
All indices run between $0$ and $N-1$. Working on individual terms we shall first derive:
\begin{equation}
\covar(\left| B_k \right|^2, \left| B_q \right|^2) = \left< \left| B_k \right|^2 \left| B_q \right|^2 \right> - \left< \left| B_k \right|^2 \right>  \left< \left| B_q \right|^2 \right> 
\end{equation}
The expansion of the product $\left| B_k \right|^2 \left| B_q \right|^2$ yields:
\begin{equation}
\left| B_k \right|^2 \left| B_q \right|^2 = \sum_{j,l,m,n} M_jM_lM_mM_n \e^{-2\imath\pi((j-l)k/N + (m-n)q/N)}
\end{equation}
The second product in the covariance can be expanded as well providing the expression:
\begin{equation}
\covar(\left| B_k \right|^2, \left| B_q \right|^2) =  \sum_{j,l,m,n} (\left< M_jM_lM_mM_n \right> - \left< M_jM_l \right> \left< M_mM_n \right>) \e^{-2\imath\pi((j-l)k/N + (m-n)q/N)}
\end{equation}
and eventually:
\begin{eqnarray}
\var(V_1^2) & =  & \sum_{k,q}\covar(\left| B_k \right|^2, \left| B_q \right|^2)  \\ \nonumber
& = &   \sum_{k,q}\sum_{j,l,m,n} (\left< M_jM_lM_mM_n \right> - \left< M_jM_l \right> \left< M_mM_n \right>) \e^{-2\imath\pi((j-l)k/N + (m-n)q/N)}
\end{eqnarray}
The $j,l,m,n$ and $k,q$ indices being independent, sums can be switched in whatever order in the above expression:
\begin{equation}
\var(V_1^2)= \sum_{j,l,m,n} \sum_{k}(\left< M_jM_lM_mM_n \right> - \left< M_jM_l \right> \left< M_mM_n \right>) \e^{-2\imath\pi(j-l)k/N} \sum_{q}\e^{-2\imath\pi(m-n)q/N}
\end{equation}
The sum over $q$ equals zero except when $m=n$ in which case it equals $N$. Thus :
\begin{equation}
\var(V_1^2)=N\sum_{j,l,m} \sum_{k}(\left< M_jM_lM_m^2\right> - \left< M_jM_l \right> \left< M_m^2 \right>) \sum_{k}\e^{-2\imath\pi(j-l)k/N}
\end{equation}
In the same way, the last sum equals zero except for $j=l$ and the expression of the variance simplifies to:
\begin{equation}
\var(V_1^2)=N^2\sum_{j,m} \left< M_j^2M_m^2\right> - \left< M_j^2 \right> \left< M_m^2 \right>
\end{equation}
Individual terms are equal to zero when $j \neq m$ as additive noises are uncorrelated yielding a still simpler expression:
\begin{equation}
\var(V_1^2)=N^2\sum_{i=0}^{N-1}\left< M_i^4\right> - \left< M_i^2 \right>^2 
\end{equation}

\subsection{Variance of $V_2^2$}
The case of $V_2^2$ is much simpler as one may easily show that the expression of the unbiased estimator is:
\begin{equation}
V_2^2 = N^2M_0^2-N^2\sigma^2 - N^2m_0
\end{equation}
As a consequence:
\begin{equation}
\var(V_2^2)=N^4\var(M_0^2)
\end{equation}

\subsection{Detector noise}
Modulated samples are considered only and are written:
\begin{equation}
M_i=m_i+b_i
\end{equation}
The variance of read-out noise is $\sigma^2$. The fourth and second moments of the gaussian distribution $M_i$ are easily calculated:
\begin{eqnarray}
\left< M_i^4 \right> & = & m_i^4 + 6m_i^2\sigma^2 + 3\sigma^4 \\ \nonumber
\left< M_i^2 \right> & = & m_i^2 + \sigma^2
\end{eqnarray}
and the variances of the estimators are:
\begin{eqnarray}
\var(V_1^2) & = & 2N^2\sigma^2 \left[ 2\sum_{i=0}^{N-1}m_i^2 + N\sigma^2 \right] \\ \nonumber
                      &	 & \\ \nonumber
\var(V_2^2) & = & 2N^4\sigma^2 \left[  2m_0^2 + \sigma^2\right]
\end{eqnarray}

\subsection{Photon noise}
Both the continuous and modulated signals contribute to photon noise. The number of photons to be considered to calculate the Poisson fluctuation per sample is therefore $c_i+m_i$ which we note $n_i$. As in \cite{tango1980}, we consider that the statistics of different samples for broadband thermal radiation are uncorrelated. Consequently, the fourth and second moments of the Poisson distribution $M_i$ are: 
\begin{eqnarray}
\left< M_i^4 \right> & = & n_i^4+6n_i^3+7n_i^2+n_i\\ \nonumber
\left< M_i^2 \right> & = & n_i^2 + n_i
\end{eqnarray}
Hence the variance of the estimators:
\begin{eqnarray}
\var(V_1^2) & = & N^2\sum_{i=0}^{N-1}n_i(4n_i^2+6n_i+1) \\ \nonumber
                      &	 & \\ \nonumber
\var(V_2^2) & = & N^4 n_0(4n_0^2+6n_0+1)
\end{eqnarray}

\subsection{Variance of estimators with filtered samples}
In practice, the fringe peak covers only a limited range of the Fourier components which are summed to compute the estimated visibility. In the derivation above, all Fourier components have been used 
thus increasing the variance. \\
Let us assume that the fringe peak has a width of $\Delta\!\sigma_0$ which corresponds to the spectral bandwidth. Whatever the nature of the noise, source photon or detector noise, its Fourier transform is a white noise and therefore the squared modulus of its Fourier transform is still a white noise. Selecting the spectral window of width $\Delta\!\sigma_0$ for the computation of the estimator amounts to reducing the variance of the noise by a factor  $\Delta\!\sigma / \Delta\!\sigma_0$, with $\Delta\!\sigma = 1/ \delta\!x$.

\subsection{Signal-to-noise ratios}
Signal-to-noise ratios can now be derived:
\begin{eqnarray}
SNR_1& = &\sqrt{ \frac{\Delta\!\sigma}{\Delta\!\sigma_0}} \times \frac{\sum_{i=0}^{N-1} m_i^2}{\sqrt{2\sigma^2 \left[ 2\sum_{i=0}^{N-1}m_i^2 + N\sigma^2 \right] + \sum_{i=0}^{N-1}n_i(4n_i^2+6n_i+1) }} \\ \nonumber
			      &     & \\ \nonumber
SNR_2& = & \sqrt{\frac{\Delta\!\sigma}{\Delta\!\sigma_0}} \times \frac{m_0^2}{\sqrt{2\sigma^2 \left[  2m_0^2 + \sigma^2\right] + n_0(4n_0^2+6n_0+1) }}
\end{eqnarray}

It is interesting to see that the SNR of the second estimator is independent of the total number of samples. The SNR of the first decreases with the number of samples. In practice, unless $N$ tends towards infinity, $V_1^2$ always has a larger SNR than $V_2^2$. Although it has been derived differently, the $V_1^2$ SNR estimate yields comparable results to the SNRs published by \cite{tango1980} and \cite{colavita1999}. An extra term however needs to be added to the result of \cite{colavita1999} and the variance due to detector noise should read $2\left< N \right>^2V^2\sigma^2 + 16\sigma^4$ instead of $16\sigma^4$ only ($\left< N \right>$ is the average total number of photons and fringe phase is constant during a sample integration). The derivation of SNRs is illustrated by two examples in the detector noise and photon noise limited regimes.

\paragraph{Detector noise limited regime}
Let us consider an interferogram with 128 samples in a 400 nm wide band centered on 2.2~$\mu$m with a sampling $\delta\!x = 0.5~\mu$m. The filter has a uniform response across the band and the spectrum is flat. We assume the amplitude of the white light fringe is $m_0=nV$ with $n$ a number of photons. We choose $V=1$. The detector noise standard deviation is set to $\sigma=n\times0.2$. With these particular values, the two SNRs are:
\begin{eqnarray}
SNR_1& = &\sqrt{ \frac{\Delta\!\sigma}{\Delta\!\sigma_0}} \times 7.63\\ \nonumber
			      &     & \\ \nonumber
SNR_2& = & \sqrt{\frac{\Delta\!\sigma}{\Delta\!\sigma_0}}\times  2.48 
\end{eqnarray}

The $V_1^2$ estimator is therefore better than $V_2^2$ for these realistic conditions which is the case as long as the number of samples remains smaller than a few thousands at high noise. 

\paragraph{Photon noise limited regime}
The same parameters are used for the photon noise limited regime. The number of uncorrelated photons per sample is set to $n=10$ yielding the SNRs:

\begin{eqnarray}
SNR_1& = &\sqrt{ \frac{\Delta\!\sigma}{\Delta\!\sigma_0}} \times 1.33\\ \nonumber
			      &     & \\ \nonumber
SNR_2& = & \sqrt{\frac{\Delta\!\sigma}{\Delta\!\sigma_0}}\times  0.54 
\end{eqnarray}

The same conclusion applies as for the detector noise limited case. $V_1^2$ is better than $V_2^2$ as long as the number of samples remains smaller than a few thousands which in practice is not required.


\begin{thebibliography}{}

\bibitem[Baldwin \& Haniff (2002)]{baldwin2002} Baldwin, J.E., Haniff, C.A. 2002, Phil. Trans. R. Soc. Lond. A., 360, 969

\bibitem[B\'erio et al. (1999)]{berio1999} B\'erio, P., Mourard, D., Bonneau, D., et al. 1999, JOSA A, 16, 872

\bibitem[Colavita (1999)]{colavita1999} Colavita, M.M. 1999, PASP, 111, 111

\bibitem[Coud\'e du Foresto et al. (1995)]{foresto1995} Coud\'e du Foresto, 
V., Perrin, G., Boccas, M. 1995, A\&A, 293, 278

\bibitem[Coud\'e du Foresto et al. (1997)]{foresto1997a} Coud\'e du Foresto, V., Mariotti, J.-M., Perrin, G. 1997, ``Direct observation of extrasolar planets with an infrared interferometer'', Science with the VLT Interferometer, Springer-Verlag, 86

\bibitem[Coud\'e du Foresto et al. (1997)]{foresto1997} Coud\'e du Foresto, 
V., Ridgway, S.T., Mariotti, J.-M. 1997, A\&AS, 121, 379

\bibitem[Davis et al. (1995)]{davis1995}ÊDavis,ÊJ., Lawson,ÊP.ÊR., Booth,ÊA.ÊJ., Tango,ÊW.ÊJ., Thorvaldson,ÊE.ÊD. 1995, MNRAS, 273, 53

\bibitem[Goodman (1985)]{goodman1985} Goodman, J. 1985, Statistical 
Optics, J. Wiley \& Sons Ed., 43

\bibitem[Hardcastle et al. (2003)]{hardcastle2003} Hardcastle,ÊM.ÊJ., Worrall,ÊD.ÊM., Kraft,ÊR.ÊP., et al. 2003,  ApJ, 593, 169

\bibitem[Kervella et al. (2003)]{kervella2004} Kervella, P., et al. 2003, A\&A, 404, 1087

\bibitem[Kotani et al. (2004)]{kotani2004} Kotani, T., Perrin, G., Woillez, J., et al., in preparation

\bibitem[Mariotti \& Ridgway (1988)]{mariotti1988} Mariotti,ÊJ.-M., Ridgway,ÊS.ÊT. 1988, A\&A, 195, 350

\bibitem[Martin et al. (2000)]{martin2000}ÊMartin,ÊF., Conan,ÊR., Tokovinin,ÊA., et al. 2000, A\&AS, 144, 39

\bibitem[Perrin et al. (2004)]{perrin2004} Perrin, G., Ridgway, S.T., Coud\'e du Foresto, V., et al. 2004, A\&A, 418, 675

\bibitem[Perrin (1997)]{perrin1997} Perrin, G. 1997, A\&AS, 121, 553

\bibitem[Tango \& Twiss (1980)]{tango1980} Tango, W.J. \& Twiss, R.Q. 1980, in Progress in Optics 17, ed. E. Wolf (Amsterdam: Noth Holland), 239

\end{thebibliography}
\end{document}